\documentstyle[prd,aps,12pt]{revtex}
\begin{document}
\draft
\title{Multidimensional geometrical model of the electrical \\
and $SU(2)$ colour charge with splitting off \\
the supplementary coordinates}
\author{ Dzhunushaliev V.D.
\thanks{E-mail address: dzhun@freenet.bishkek.su}}
\address {Department of the Theoretical Physics \\
          Kyrgyz State National University, Bishkek, 720024}
\maketitle
\begin{abstract}
The geometrical model of an electrical charge is proposed. This model has 
the ''nake'' charge shunted with {\em ''fur - coat''} consisting of 
virtual wormholes. The 5D wormhole solution in the Kaluza - 
Klein's theory is the ''nake'' charge. The splitting off the supplementary 
coordinates happens on the two spheres (null surfaces) bounding this 
5D wormhole. This allows to sew two  Reissner - Nordstr\"om's black holes to 
it on both sides. Virtual wormholes entrap a part of the electrical force 
lines outcoming from ''nake'' charge. This effect can essentially 
decrease the charge visible at infinity up to real relation $m^2<e^2$. The 
analogical construction for colour $SU(2)$ gauge charge is made.
\end{abstract}

\section{Introduction}
The existence of an intrinsic electrons structure may be a  
radical remedy for singularity in quantum field theory. In due time 
J.Wheeler \cite{wh1} - \cite{wh3} had proposed a model ''charge without 
charge''. Its idea lies in the fact that the wormhole (WH), entrapping the 
electrical force lines, is an electrical charge for external observer. The 
charge sign depends upon how WH part is observed: with entering or outgoing 
force lines. In first case this is a negative charge and in the second case 
it is a positive charge. 
\par
This J.Wheeler model is 4 dimensional. On the other hand, the 5 dimensional 
Kaluza - Klein's theory has the wonderful properties: In this theory the 
''electrical'' field is a pure geometrical object. These two viewpoints on 
the electrical field has been joined in Ref.\cite{dzh1}. The $5D$ solution 
in the Kaluza - Klein's theory is founded in this work, that is Lorentzian 
WH, and it is bounded by two null surfaces. This WH can be sewed with 2 
stationary Reissner - Nordstr\"om's solutions. Thus, the obtaining object is 
a composite WH with following properties:\begin{enumerate}
\item
The intrinsic part is the $5D$ Lorentzian WH.
\item
Two exterior  parts are the 4D stationary asymptotical flat regions as a 
stationary Reissner - Nordstr\"om's solution $(r>r_+, r_+ = m + \sqrt{m^2 - 
e^2}$ is event horizon, $m$ and $e$ are mass and charge of the black hole, 
respectively). One region contains the entering force lines into $5D$ WH and 
the another region contains the outcoming electrical force lines from $5D$ 
WH.\item
The sewing of the intrinsic region with each exterior region 
happens on the event horizons $r_+$, that guarantees the multidimensional 
world nonobservability for exterior observer.
\item
The $5D$ supplementary coordinate splits off on the surface sewing. This 
results in that the  $G_{5t}$ component of the intrinsic metrical 
tensor generates the $4D$ electrical field in the exterior regions.
\end{enumerate}

\section{The composite wormhole as an electrical charge}

At first we remind the results achieved in Ref.\cite{dzh1}.  
$5D$ metric has the following WH-like view:
\begin{equation}
ds^2 = e^{2\nu (r)}dt^2 - e^{2\psi (r)} \left (d\chi - \omega (r)dt
\right )^2 - dr^2 - a^2(r)\left (d\theta ^2 + \sin ^2\theta d\varphi ^2
\right ),
\label{11}
\end{equation}
where $\chi$ is 5 supplementary coordinate; $r, \theta ,\varphi$ 
are 3D polar coordinates; $t$ is time. Corresponding $5D$ Einstein's  
equations have the following solution:
\begin{eqnarray}
a^2 & = & r^2_0 + r^2,
\label{12a}\\
e^{-2\psi}= e^{2\nu} & = & {{2r_0}\over q} {{r^2_0 + r^2}
\over{r^2_0 - r^2}},
\label{12b}\\
\omega &= & {{4r^2_0}\over q} {r\over{r^2_0 - r^2}},
\label{12c}
\end{eqnarray}
where $r_0$ is a throat of given wormhole; $q$ is a 5D 
''electrical'' charge. It is easy to see that the time component of 
metrical tensor $G_{tt} (r=\pm r_0)=0$. This indicates that there is the 
null surface, as on its $ds^2=0$. The sewing of the $5D$ and $4D$ physical 
quantities happens in the following manner:
\begin{eqnarray}
e^{2\nu _0} - \omega ^2_0 e^{-2\nu _0} = G_{tt}\left 
(\pm r_0\right ) = g_{tt}\left (r_+\right ) = 0,
\label{13a}\\
r^2_0 = G_{\theta\theta}(\pm r_0) = g_{\theta\theta}(r_+) = r^2_+,
\label{13b}
\end{eqnarray}
where $G$ and $g$ are $5D$ and $4D$ metrical tensors, respectively. 
$r_+$ is event horizon for Reissner - Nordstr\"om's 
solution. The quantity marked by $(0)$ sign are taken by $r=\pm r_0$. The 
sewing $G_{\chi t}$ and $4D$ electrical field happens in the following 
manner \cite{dzh1}:
\begin{equation}
{q\over{2r^2_0}} = {e\over{r^2_+}}.
\label{14}
\end{equation}
The Reissner - Nordstr\"om's condition $m^2>e^2$ is the basic 
cause obstructing to interpret such composite WH as an electron model. 
\par
The quantum gravity confirms that in microscopic scale the metric 
fluctuations are so large that the topological fluctuations - wormholes 
(handles) appear in spacetime. Such fluctuations have  to spring up in 
regions with very strong gravitational field. It just happens near  
the event horizon surface $r=r_+$ of the sufficiently small black hole.
\par
In a suggested model it is assumed that the virtual wormholes (VWH) 
arise between two 4D regions with incoming and outcoming force lines and the 
strong gravitational field near 2 surfaces $r=r_+$. The appearance of this 
VWH leads that they entrap the part of $4D$ electrical force lines. If the 
total cross size of all VWH is of the same order as the cross size of the 
$5D$ WH, then the VWH entrap almost all electrical force lines outcoming 
from $5D$ WH. In this case the exterior observer can detect electrical 
charge with the real relation between mass and charge $m^2 < e^2$.
\par
Thus, in the suggested model the real 
electrical charge consists of ''nake'' electrical charge ($5D$ WH) and 
VWH {\em ''fur - coat''} dressed on it. Such {\em ''fur - coat''} 
essentially decreases the electrical charge visible at infinity, i.e. in 
fact makes its renormalization. This model cannot yet describe the charge 
quantization and $\hbar /2$ spin of electron. In the first case it is 
necessary to have the quantum field theory of gravity which we do not yet 
have. The second case is considered below.

\section{$SU(2)$ composite wormhole}

The above suggested $5D$ model of the electrical charge can be 
generalized to the $SU(2)$ colour charge case. We consider gravity on the 
total space of a bundle with base being $4D$ Einstein's spacetime and fibre 
being $SU(2)$ gauge group. In short, we cite results achieved in 
Ref.\cite{dzh2}. In this case the gravity equation looks as a follows:
\begin{eqnarray}
R_{A\mu} -{1\over 2}G_{A\mu }R = 0,
\label{31}\\
R^{a}_{a} = R^{4}_{4} + R^{5}_{5} + R^{6}_{6} = 0,
\label{32}
\end{eqnarray}
where $R_{A\mu }$ is Ricci tensor and $R$ is Ricci scalar; 
$A = 0,1,2,\ldots ,6$ is multidimensional index on the total bundle space; 
$\mu = 0,1,2,3$ 
 is spacetime index on the base of bundle; $a=4,5,6$ is the 
coordinate index on the fibre of bundle. The fibre is a symmetrical space 
hence the metric on the fibre has only one degree of freedom (conformal 
factor of the fibre metric), from here leads the Eq.($\ref{32}$). In our 
case the multidimensional metric has the following form:
\begin{eqnarray}
ds^{2} = e^{2\nu (r)}dt^{2} & - & r^{2}_{0}e^{2\psi 
(r)}\sum^{6}_{a=4}\left (\sigma ^{a} - A^{a}_{\mu }(r)
dx^{\mu }\right )^{2} -
\nonumber\\
& & dr^{2} - a^{2}(r)\left (d\theta ^{2} + \sin ^{2}\theta 
d\varphi ^2\right ).
\label{33}
\end{eqnarray}
The "potentials" $A^{a}_{\mu }$ have the following monopole-like form:
\begin{eqnarray}
A^{a}_{i} & = & \frac{\epsilon ^{a-3}_{ij}x^j}{r^2}[f(r) + 1],
\label{41}\\
A^{a}_{t} & = & \frac{x^a}{r^2} v(r).
\label{42}
\end{eqnarray}
here $i,j=1,2,3$ are space indexes. We examine the most interesting case 
when $f(r)=0$ and the  linear dimensions  of fibre $r_{0}$  are  vastly  
smaller  than  the  space dimension $a_{0}$ and "charge" $q$ is sufficiently 
large:
\begin{equation}
\left ({q\over a_{0}}\right )^{1/2} \gg 
\left ({a_0\over r_{0}}\right )^2 \gg  1,
\label{16}
\end{equation}
 where $a_{0}=a(r=0)$ is the throat of the WH, $q$  defines the 
$A^a_{t}$ potential:
\begin{equation}
v'  ={q\over r_{0}a^{2}} e^{\nu -5\psi },
\label{6}
\end{equation}
In this case we can achieve the approximate solution of the 
(\ref{31})-(\ref{32}) system in the following form -\cite{dzh2}:
\begin{eqnarray}
\nu & = & -3\psi,
\label{71}\\
a^2 & = & a^2_0 + r^2,
\label{72}\\
e^{-{4\over 3}\nu } & = & {q\over 2a_0}
\cos \left (\sqrt {8\over 3}\arctan {r\over a_0} 
     \right ),
\label{73}\\
v & = & \sqrt 6{a_{0}\over r_0 q}
\tan \left (\sqrt {8\over 3} \arctan {r\over a_0}
     \right ).
\label{74}
\end{eqnarray}
It is easy to shown that $ds^2=0$ by $r=\pm a_0$, hence $r=\pm a_0$ surfaces 
are null surfaces.
\par
Analogously to above - mentioned $5D$ case we can construct the 
composite WH consisting of intrinsic 7D WH (''nake'' $SU(2)$ colour 
charge) and two $4D$ Einstein - Yang - Mills black hole filled by 
chromoelectrical $A^a_t$ field and pasting to $r = \pm r_0$ null surfaces. 
Also such $SU(2)$ composite WH can be dressed a {\em ''fur - coat''} 
consisting from VWH, that naturally leads to decreasing the $SU(2)$ field 
far from the centre. From the viewpoint of a far observer such composite 
$SU(2)$ WH is the $SU(2)$ colour charge with nonzero rest - mass and very 
weak $SU(2)$ field.
\par
We remind the following result \cite{per}-\cite{sal}. Let $G$  group 
 be  the fibre of principal  bundle.  Then  there  is  the  one-to-one
correspondence between $G$-invariant metrics on the  total  space 
${\cal X}$ and the triples $(g_{\mu \nu }, A^{a}_{\mu }, h\gamma _{ab})$. 
Where $g_{\mu \nu }$ is Einstein's pseudo  -
Riemannian metric; $A^{a}_{\mu }$ is gauge field  of  the $G$  group; 
$h\gamma _{ab}$  is symmetric metric on the fibre. These remarks allow to 
continue the multidimensional ''potentials'' $A^a_{\mu}$ into 4D regions as 
a true gauge potential of Yang - Mills field.

\section{Discussion}

There are 3 basic problems relating to above constructed WH with electrical 
(colour) charge:
\begin{enumerate}
\item
What parts of the force lines outcoming from $5D (7D)$ WH are entrapped 
by VWH {\em ''fur - coat''}?
\item
What spin has such composite wormhole?
\item
What is the mechanism of supplementary coordinates splitting off?
\end{enumerate}
The first question can be solved only in the framework of future 
quantum gravity. It is possible that in the second case $\hbar /2$ spin may 
be obtained using the mechanism similar to ''spin isospin'' phenomenon. 
 The third question was discussed in 
Ref.\cite{dzh3}. In this article the viewpoint is proposed, according 
to which some quantum transition in quantum gravity (splitting off the 
supplementary coordinates, changing of metric signature, Universe birth) is 
connected with changing the algorithmic complexity of corresponding gravity 
objects. This viewpoint is based on A.N. Kolmogorov's theory, according to 
which the probability of appearance of some objects can be connected with 
algorithm length describing this object.

\end{document}